\documentclass[10pt]{article}
\usepackage{graphicx, array, blindtext}
\usepackage[T1]{fontenc}
\usepackage{authblk}
\usepackage{amssymb,amsmath}
\usepackage{anyfontsize}
\usepackage{amsmath}

\begin{document}			

\title{\Large{A Matter with an effective EoS interacting with a tachynic field in an accelerating Universe}}
\author[1,2]{\large{Martiros Khurshudyan\thanks{martiros.khurshudyan@nano.cnr.it, khurshudyan@yandex.ru}}}

\affil[1]{\footnotesize{CNR NANO Research Center S3, Via Campi 213a, 41125 Modena MO}}
\affil[2]{\footnotesize{Dipartimento di Scienze Fisiche, Informatiche e Matematiche, Universita degli Studi di Modena e Reggio Emilia, Modena, Italy}}
\renewcommand\Authands{ and }

  \maketitle

\begin{abstract}
We consider a fluid described by a parameterized EoS of the general form $P=(\gamma-1)\rho+p_{0}+\omega_{H}H+\omega_{H2}H^{2}+\omega_{dH}\dot{H}$ \cite{Ren}, where $p_{0}$, $\omega_{H}$, $\omega_{H2}$ and $\omega_{dH}$ are free parameters of the model, interacting with a Tachyonic field with a relativistic Lagrangian $L_{TF}=-V(\phi)\sqrt{1-\partial_{i}\phi\partial^{i}\phi}$. The acceleration of the Universe described by a scale factor $a(t)=t^{n}, (n>1)$. Under consideration of different forms of interaction the field $\phi$ and the potential $V(\phi)$ are recovered and graphical analysis performed. For illustration purposes we fixed values of parameters of the models to provide $V \rightarrow 0$ for later stages of evolution, when $t \rightarrow \infty$.
\end{abstract}
\newpage
\section*{\large{Introduction}}

The observations of high redshift type SNIa supernovae \cite{Riess} reveal the speeding up expansion of our universe. The surveys of clusters of galaxies show that the density of matter is very much less than critical density \cite{Pope}, observations of Cosmic Microwave Background anisotropies indicate that the universe is flat and the total energy density is very close to the critical $\Omega_{\small{tot}} \simeq1$ \cite{Spergel}. Finding the theoretical explanation of cosmic acceleration has been one of the central problems of modern cosmology and theoretical physics. In order to explain experimental data concerning to the nature of the accelerated
expansion of the Universe a huge number of hypothesis were proposed. For instance,
in General Relativity framework, the desirable result could be achieved by so-called dark energy: an exotic and mysterious component of the Universe, with negative pressure
(we thought that the energy density is always positive) and with negative EoS parameter
$\omega<0$\footnote{however the negative energy density is also an interesting subject of
investigation and were considered by several authors including Stephen Hawking.}.  Dark energy occupies about 73$\% $ of the energy of our universe, other componet, Dark matter, about 23$\%$, and usual baryonic matter accupy about 4$\%$. The simplest model
for a dark energy is a cosmological constant $\omega_{\Lambda}=-1$ introduced by Einstein, but  with cosmological constant we faced with two problems i.e. absence of a fundamental mechanism which sets the cosmological constant zero or very small value the problem known as fine-tuning problem, because in the freamwork of quantum field theory, the expectation value of vacuum energy is 123 order of magnitude larger than the observed value \cite{Steinhardt}. The second problem known as cosmological coincidence problem, which asks why are we living in an epoch in which the densities of dark energy and matter\footnote{The other dark component is a dark matter, which also could have a role in the acceleration of the expansion of the Universe, but we should argue, that unfortunately we have not enough information about that and we have to consider models on phenomenological level with hope to find some understanding, which is completely hard scientific research.} are comparable? Alternative models of dark energy suggest a dynamical form of dark energy, which at least in an effective level, can originate from a variable cosmological constant \cite{Sola}, or from various fields, such is a canonical scalar field \cite{Ratra} (quintessence), a phantom field, that is a scalar field with a negative sign of the kinetic term \cite{Caldwell}, \cite{Caldwell1}, or the combination of quintessence and phantom in a unified model named quintom \cite{Feng} and could alleviate these problems. Finally, an interesting attempt to probe the nature of dark energy according to some basic quantum gravitational principles are the holographic dark energy paradigm \cite{Hsu} and agegraphic dark energy models \cite{Cai}. In order to explain the current accelerated expansion without introducing dark energy, one may use a simple generalized version of the so-called teleparallel gravity \cite{Einstein}, namely $F(T)$ theory. It is a generalization of the teleparallel gravity by replacing the so-called torsion scalar $T$ with $F(T)$. $TG$ was originally developed by Einstein in an attempt of unifying gravity and electromagnetism. $F(T)$ gravity is not locally Lorentz invariant and appear to harbor extra degrees of freedom not present in general relativity \cite{Li}.  Although teleparallel gravity is not an alternative to general relativity (they are dynamically equivalent), but its different formulation allows one to say: gravity is not due to curvature, but to torsion. In other word, using tetrad fields and curvature-less Weitzenbock connection instead of torsion-less Levi-Civita connection in standard general relativity. Modifications of the Hilbert-Einstein action by introducing different functions of the Ricci scalar $R$ have been systematically explored, the so-called $F(R)$ gravity models, which reconstruction has been developed \cite{Nojiri}-\cite{Elizalde} and, for instance, modified Gauss-Bonnet gravity, that is, a function of the GB invariant \cite{Nojiri2} are other attemptes to explain acceleration without DE. By the way, the field equations for the $F(T)$ gravity are very different from those for $f(R)$ gravity, as they are second order rather than fourth order. \\ \\
Futumore, since no known symmetry in nature prevents or suppresses a nonminimal coupling between dark energy and dark matter, there may exist interactions between the two components. At the same time, from observation side, no piece of evidence has been so far presented against such interactions. Indeed, possible interactions between the two dark components have been discussed in recent years. It is found that a suitable interaction can help to alleviate the coincidance problem. Different interacting models of dark energy have been investigated. For instance, the interacting Chaplygin gas allows the universe to cross the phantom divide: the transition from $\omega>−1$ to $\omega<−1$, which is not permissible in pure Chaplygin gas models. \\
The model with interaction between dark energy and dark matter describes by the Friedmann equation 
\begin{equation}\label{eq: Fridmman vlambda}
H^{2}=\frac{\dot{a}^{2}}{a^{2}}=\frac{\rho_{\small{tot}}}{3},
\end{equation}
as the reduced result of the field equations
\begin{equation}\label{eq:Einstein eq}
R^{\mu\nu}-\frac{1}{2}g^{\mu\nu}R^{\alpha}_{\alpha}=T^{\mu\nu},
\end{equation}
with FRW metric (the metric of a spatially flat homogeneous and isotropic universe)
\begin{equation}\label{eq:FRW metric}
ds^{2}=dt^{2}-a(t)^{2}\left(dr^{2}+r^{2}d\theta^{2}+r^{2}\sin^{2}\theta d\phi^{2}\right).
\end{equation}
and two conservation laws
\begin{equation}\label{eq:DM}
\dot{\rho}_{\small{DM}}+3\frac{\dot{a}}{a}(\rho_{\small{DM}}+P_{\small{DM}})=Q.
\end{equation}
\begin{equation}\label{eq:DE}
\dot{\rho}_{\small{DE}}+3\frac{\dot{a}}{a}(\rho_{\small{DE}}+P_{\small{DE}})=-Q,
\end{equation}
where $H=\frac{\dot{a}}{a}$ is Hubble parameter, $a$ is a scale factor and $Q$ denotes the phenomenological interaction term. Last two equations could be understood as follows: as there is an interaction between components there is not energy conservation for the components separately, but for the whole mixture the energy conservation is hold. This approach could work as long as we are working without knowing the actual nature of the dark energy and dark matter as well as about the nature of the interaction. This approach at least from mathematical point of view is correct. The forms of interaction term considered in literature very often are of the following forms: $Q=3Hb\rho_{dm}$, $Q=3Hb\rho_{\small{de}}$, $Q=3Hb\rho_{\small{tot}}$, where $b$ is a coupling constant and positive $b$ means that dark energy decays into dark matter, while negative $b$ means dark matter decays into dark energy. From thermodynamical view, it is argued that the second law of thermodynamics strongly favors dark energy decays into dark matter. However it was found that the observations may favor the decaying of dark matter into dark energy. Other forms for interaction term considered in literature are $Q=\gamma\dot{\rho}_{\small{dm}}$, $Q=\gamma\dot{\rho}_{\small{de}}$, $Q=\gamma\dot{\rho}_{\small{tot}}$, $Q=3Hb\gamma\rho_{i}+\gamma\dot{\rho_{i}}$, where $i=\{dm,de,tot\}$. These kind of interactions are either positive or negative and can not change sign. However, recently by using a model independent metod to deal with the observational data Cai and Su found that the sign of interaction $Q$ in the dark sector changed in the redshift range of $0.45 \lesssim z \lesssim 0.9$.  Hereafter, a sign-changeable interaction \cite{Hao},\cite{Hao2} were introduced 
\begin{equation}\label{eq:signcinteraction}
Q=q(\alpha\dot{\rho}+3\beta H\rho).
\end{equation}
where $\alpha$ and $\beta$ are dimensionless constants, the energy density $\rho$ could be $\rho_{m}$, $\rho_{\small{de}}$, $\rho_{tot}$. $q$ is the deceleration parameter
\begin{equation}\label{eq:decparameter}
q=-\frac{1}{H^{2}} \frac{\ddot{a}}{a}=-1-\frac{\dot{H}}{H^{2}}.
\end{equation}
This new type of interaction, where deceleration parameter $q$ is a key ingredient makes this type of interactions different from the ones considered in literature and presented above, because it can change its sign when our universe changes from deceleration $q>0$ to acceleration $q<0$. $\alpha \dot{\rho}$ is introduced from the dimensional point of view. We would like also to stress a fact, that by this way we import a more information about the geometry of the Universe into the interaction term. \\
In this article we will consider a mixture of a scalar field given by relativistic Lagrangian and known as Tachyonic field \cite{Sen}.
\begin{equation}\label{eq:TF}
L=-V(\phi)\sqrt{1-\partial_{\mu}\phi\partial^{\mu}\phi}.
\end{equation}
The stress energy tensor
\begin{equation}\label{eq:stresstensor}
T^{\mu\nu}=\frac{\partial L}{\partial(\partial_{\mu}\phi)}\partial^{\nu}-g^{\mu\nu}L.
\end{equation}
gives the energy density  and pressure as
\begin{equation}\label{eq:TFdensity}
\rho_{\small{TF}}=\frac{V(\phi)}{\sqrt{1-\dot{\phi}^{2}}}.
\end{equation}
and
\begin{equation}\label{eq:TFpressure}
P_{\small{TF}}=-V(\phi)\sqrt{1-\dot{\phi}^{2}}.
\end{equation}
and a fluid with modified equation of state \cite{Ren}
\begin{equation}\label{eq:modEOS}
P=(\gamma-1)\rho+p_{0}+\omega_{H}H + \omega_{H2}H^{2}+\omega_{dH}\dot{H}.
\end{equation}
which with $p_{0}=\omega_{H}=\omega_{H2}=\omega_{dH}=0$ and $\gamma-1=\omega_{b}$ will reduce to the EoS of a barotropic fluid. We would like to refer our readers to the series of works, were similar concepts were developed and considered \cite{Sergey}. In \cite{Ujjal} an interaction between barotropic fluid and Tachyonic scalar field of the $Q=3Hb\rho_{m}$ form was considered and field $\phi$ as well as potential $V(\phi)$ were obtained. We will follow to the line as in \cite{Ujjal}, but with the new fluid and we will consider sign-changeable interaction. The mixture of our consideration describes by $\rho_{\small{tot}}$ and $P_{\small{tot}}$ given by
\begin{equation}\label{eq:mixture energy}
\rho_{ \small{tot}} = \rho_{ \small{DM}}+\rho_{ \small{DE}}.
\end{equation}
and
\begin{equation}\label{eq:mixture presure}
P_{ \small{tot}} = P_{ \small{DM}}+P_{ \small{DE}}.
\end{equation}
Statefinder diagnostic for the model is also presented after proper introduction to the stafinder dignostic tool. Accelerating Universe will be described by a $a(t)=t^{n}$, with $n>1$, scale factor. Power-law cosmology, where scale factor is a power of the cosmological time i.e our case, proves to be very good phenomenological description of the universe evolution, because it can describe the radation epoch, the dark matter epoch and accelerating epoch according to the value of the exponent and seems suported by the observational data. 
\\ \\
Paper organized as follow: Basic ideas and motivation, with field equations, space-time metric, interaction forms and phenomenology, descriptions of the dark energy and matter, with a phenomenological coupling are given in introduction section. Next, we present problem solving strategy, then for each model we found $\phi$, potential $V(\phi)$ as well as analyse profile of $\omega_{tot}$. Some conclusion is given at the end of work.

\section{Non interacting case}
Pressure for our phenomenological fluid with a fixed scale factor of $t^{n}$ could be writen as follow
\begin{equation}\label{eq:pscale}
P=(\gamma-1)+P_{0}+n\omega_{H}t^{-1}+n^{2}\omega_{H2}t^{-2}-n\omega_{dH}t^{-2}.
\end{equation}
Absence an interaction between components of the mixture means that they evolve separately and (\ref{eq:DM}) and (\ref{eq:DE}) will take the forms
\begin{equation}\label{eq:noninteqm}
\dot{\rho}_{m}+3H(\rho_{m}+P_{m})=0.
\end{equation}
and
\begin{equation}\label{eq:noninteqG}
\dot{\rho}_{\small{TF}}+3H(\rho_{\small{TF}}+P_{\small{TF}})=0.
\end{equation}
A solution of (\ref{eq:noninteqm}) accounting (\ref{eq:pscale}) reads as
\begin{equation}\label{eq:rhomnonint}
\rho_{m}=\rho_{0}e^{-3n\gamma}-3nE.
\end{equation}
For dark energy density we will get
\begin{equation}\label{eq:denonintdensity}
\rho_{\small{TF}}=3\frac{n^{2}}{t^{2}}-\rho_{0}e^{-3n\gamma}+3nE.
\end{equation}
Solving (\ref{eq:noninteqG}) we obtain $P_{\small{TF}}$
\begin{equation}\label{eq:denonintpressure}
P_{\small{TF}}=-\frac{p_{0}}{\gamma}-(\gamma-1)\rho_{0}t^{-3n\gamma}+\frac{n}{t^{2}}\ast B -A.
\end{equation}
and taking into account that
\begin{equation}\label{eq:TFEOSpar}
\omega_{\small{TF}}=\frac{P_{\small{TF}}}{\rho_{\small{TF}}}=-(1-\dot{\phi}^{2}).
\end{equation}
for a field $\phi$ and potential $V(\phi)$ we obtain
\begin{equation}\label{eq:fieldnonint}
\phi=\int{\sqrt{1+\omega_{\small{tot}}}~dt}.
\end{equation}
$$
V(\phi)=\sqrt{\frac{p_{0}}{\gamma} -\rho_{0}t^{-3n\gamma}+\frac{3n^{2}}{t^{2}}D +C}\times$$
\begin{equation}\label{eq:potnonint}
\times\sqrt{\frac{p_{0}}{\gamma}+\rho_{0}(\gamma -1)t^{-3n\gamma}-\frac{n}{t^{2}}\ast B+A}.
\end{equation}
where $A$, $B$, $C$, $D$ and $E$ are
$$A=C-n^{2}\left( \frac{3\omega_{H}}{t-3nt\gamma}+\frac{6-9n\gamma+3\omega_{dH}+2\omega_{\small{H2}}}{t^{2}(-2+3n\gamma)}\right),$$
$$B=2+\frac{2\omega_{\small{dH}}}{2-3n\gamma}+\frac{t\omega_{H}}{-1+3n\gamma},$$
$$C=\frac{3n^{3}\omega_{\small{H2}}}{t^{2}(-2+3n\gamma)},$$
$$D=1+\frac{\omega_{\small{dH}}}{2-3n\gamma} - \frac{t\omega_{\small{H}}}{1-3n\gamma},$$
$$E=\frac{p_{0}}{3\gamma n}-\frac{n\omega_{H}}{t-3n\gamma t}+
\frac{n(n\omega_{\small{H2}}-\omega_{\small{dH}})}{(-2+3n\gamma)t^{2}}.$$
\begin{figure}
\centering
\includegraphics[width=70 mm]{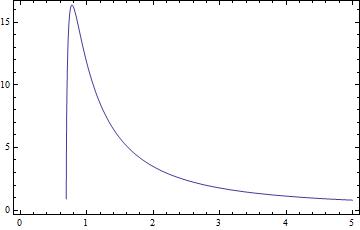}
\caption{
\small{The variation of $V$ against t: Non interacting case, $n=2$, $\rho_{0}=1$, $p_{0}=0$, $\omega_{H}=1.5$, $\omega_{H2}=0.5$ and $\omega_{dH}=1$.}} 
\label{fig:potnint}
\end{figure}
Analysis shows that, when free parameter $p_{0}=0$, then $V\rightarrow0$ with time thus retaining the original property of the tachyon potential. Having a tiny non zero value for $p_{0}$ it is always possible to obtain $V\rightarrow 0$ with time. For fixing a reasonable diapason of values for parameters obtained results should be compared with observational data, which will be done in forthcoming articles. In this case $\omega_{\small{tot}}$ indicates quintessence-like behavior during whole evolution of the Universe: from early epoch to late stage.
\section{Interacting case}\label{s:interaction}
In this section we will consider different forms of interaction intensively considered in literature: $Q=3Hb\rho$, $Q=\gamma\dot{\rho}$ and $Q=q(\alpha\dot{\rho}+3\beta H \rho)$ known as sign-changeable interaction. In all types of interaction under consideration $\rho$ could be $\rho_{m}$, $\rho_{\small{DE}}$ or $\rho_{\small{tot}}$.
\subsection{$Q=3Hb\rho_{m}$}\label{ss:rho}
\begin{figure}
\centering
\includegraphics[width=70mm]{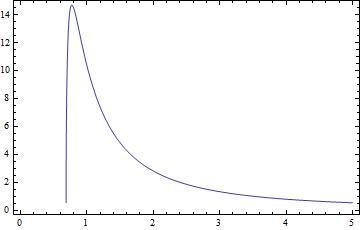}
\caption{%
\small{The variation of $V$ against t, Interaction: $Q=3Hb\rho_{m}$,
Parameters: $n=2$, $\rho_{0}=1$, $p_{0}=0$, $\omega_{H}=0.5$, $\omega_{H2}=0.5$ and $\omega_{dH}=1$, $b=0.05$.}} 
\label{fig:potintrho}
\end{figure}

\begin{figure}
\centering
\includegraphics[width=70 mm]{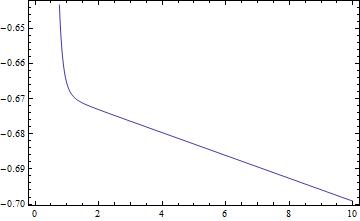}
\caption{%
The variation of $\omega_{\small{tot}}$ against t, Interaction: $Q=3Hb\rho_{m}$,
Parameters: $n=2$, $\rho_{0}=1$, $p_{0}=0$, $\omega_{H}=0.5$, $\omega_{H2}=0.5$ and $\omega_{dH}=1$, $b=0.05$.} 
\label{fig:omegatotrho}
\end{figure}

With interaction term $Q=3Hb\rho_{m}$ the solution of (\ref{eq:inteqm}) reads as
\begin{equation}\label{eq:intrhomatter}
\rho_{m}=\rho_{0}t^{3n(b-\gamma)}+\frac{p_{0}}{b-\gamma}+3n^{2}A_{1}.
\end{equation}
where $A_{1}$ is
$$A_{1}=\frac{\omega_{H}}{t+3nt(b-\gamma)}-\frac{\omega_{dH}-n\omega_{H2}}{t^{2}(2+3n(b-\gamma))}.$$
Energy density and pressure read as
\begin{equation}\label{eq:intderho}
\rho_{\small{TF}}=\frac{3n^{2}}{t^{2}}-\rho_{m}.
\end{equation}
and
\begin{equation}\label{eq:intdepressure}
P_{\small{TF}}=\frac{-Q-\dot{\rho}_{\small{TF}}}{3H}-\rho_{\small{TF}}.
\end{equation}
For a field we can use (\ref{eq:fieldnonint}) to obtain its explicit form. For the potential
we have $V(\phi)=\sqrt{-\rho_{\small{TF}}P_{\small{TF}}}$, where the minus will not make any
problem, because $P_{\small{TF}}<0$. The graphical analysis of $V(\phi)$ and $\omega_{\small{tot}}$ are presented in (Fig. \ref{fig:potintrho})
and (Fig.\ref{fig:omegatotrho}). From (Fig.\ref{fig:omegatotrho}) it is clear, that in this case $\omega_{\small{tot}}>-1$
and indicates quintessence-like behavior. $V\rightarrow0$ with time could be obtained as in the previous case.
\subsection{$Q=b\dot{\rho}_{m}$}\label{ss:dotrho}
In this section we will use another form of interaction, which is proportional to $\dot{\rho}_{m}$.
Presence of this kind of interaction for a matted density $\rho_{m}$ gives us
\begin{equation}\label{}
\rho_{m}=\rho_{0}t^{\frac{3n\gamma}{b-1}}-\frac{p_{0}}{\gamma}-\frac{3n^{2}}{t^{2}}A_{2},
\end{equation}
where $A_{2}$ is
$$A_{2}=\frac{t\omega_{\small{H}}}{-1+b+3n\gamma}-\frac{\omega_{\small{dH}}-n\omega_{\small{H2}}}{-2+2b+3n\gamma}.$$
Following the same mathematical line as in above sections, we can recover $\rho_{\small{TF}}$, $P_{\small{TF}}$,
$\omega_{\small{TF}}$ which gives the field $\phi=\int{\sqrt{1+\omega_{\small{TF}}}}~dt$as well as a potential $V(\phi)$ reads as
$$V(\phi)=\sqrt{\frac{p_{0}}{\gamma}-\rho_{0}t^{\frac{3n\gamma}{b-1}}+\frac{3n^{2}}{t^{2}}B_{2}}\times$$\begin{equation}
\label{eq:potintdrho}
\times\sqrt{\frac{p_{0}}{\gamma}-\frac{\rho_{0}}{-1+b}(-1+b+\gamma)t^{\frac{3n\gamma}{-1+b}}-\frac{n}{t^{2}}C_{2}+\frac{n^{2}}{t^{2}}D_{2}}.
\end{equation}
where $B_{2}$, $C_{2}$ and $D_{2}$ are
$$B_{2}=1+\frac{\omega_{\small{dH}}+n\omega_{\small{H2}}}{-2+2b+3n\gamma}+\frac{t\omega_{\small{H}}}{-1+b+3n\gamma},$$
$$C_{2}=2-\frac{2\omega_{\small{dH}}}{-2+2b+3n\gamma}+\frac{t\omega_{\small{H}}}{-1+b+3n\gamma},$$
$$D_{2}=3+\frac{3t\omega_{\small{H}}}{-1+b+3n\gamma}-\frac{3\omega_{\small{dH}}+(2-3n)\omega_{\small{H2}}}{-2+2b+3n\gamma}.$$
\begin{figure}
\centering
\includegraphics[width=70 mm]{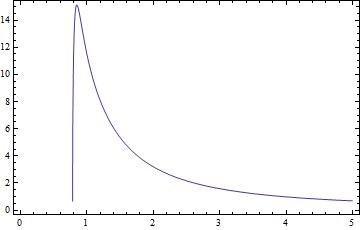}
\caption{%
\small{The variation of $V$ against t, Interaction:  $Q=b\dot{\rho}_{m}$,
Parameters: $n=2$, $\rho_{0}=1$, $p_{0}=0$, $\gamma=2$, $\omega_{H}=1.5$, $\omega_{H2}=0.5$ and $\omega_{dH}=1$, $b=0.05$.}} 
\label{fig:potintdrho}
\end{figure}
\\For this type of interaction we were able to see that, when free parameter $p_{0}=0$, then $V\rightarrow0$ with time thus retaining the original property of the tachyon potential. Having a tiny non zero value for $p_{0}$ it is always possible to obtain $V\rightarrow 0$.
This model with $\omega_{\small{tot}}>-1$ indicates quintessence-like behavior.
\subsection{$Q=q(\alpha\dot{\rho}_{m}+3\beta H\rho_{m})$}\label{ss:dotrho}
\begin{figure}
\centering
\includegraphics[width=70 mm]{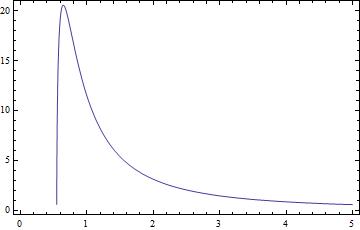}
\caption{%
\small{The variation of $V$ against t, Interaction: $Q=q(\alpha\dot{\rho}_{m}+3\beta H\rho_{m})$,
Parameters: $n=2$, $\rho_{0}=1$, $p_{0}=0$, $\omega_{H}=0.5$, $\omega_{H2}=1$ and $\omega_{dH}=1$, $\alpha=1.2$, $\beta=0.4$.}} 
\label{fig:potintschange}
\end{figure}

\begin{figure}
\centering
\includegraphics[width=70 mm]{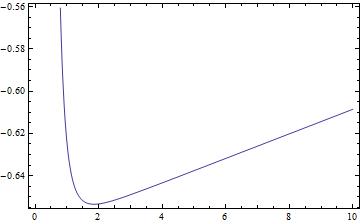}
\caption{%
\small{The variation of $\omega_{\small{tot}}$ against t, Interaction: $Q=q(\alpha\dot{\rho}_{m}+3\beta H\rho_{m})$,
Parameters:  $n=2$, $\rho_{0}=1$, $p_{0}=0$, $\omega_{H}=0.5$, $\omega_{H2}=1$ and $\omega_{dH}=1$, $\alpha=1.2$, $\beta=0.4$.}} 
\label{fig:omegatotschange}
\end{figure}

Investigation of the model in case of sign-changeable interaction reveals the following behavior: The solution of (\ref{eq:inteqm}) gives us the following result for the energy density of a matter
\begin{equation}\label{eq:rhomcint}
\rho_{m}=\rho_{0}t^{-\frac{3n(\beta(-1+n)+n\gamma)}{\alpha(-1+n)+n}}-\frac{np_{0}}{\beta(-1+n)+n\gamma}+\frac{3n^{2}}{t^{2}}A_{3}.
\end{equation}
where $A_{3}$ is
$$A_{3}=\frac{nt\omega_{\small{H}}}{\alpha(-1+n)+n(1-3\beta(-1+n)-3n\gamma)}-$$ $$-\frac{3n(\omega_{\small{dH}}-n\omega_{\small{H2}})}{2\alpha(-1+n)+n(2-3\beta(-1+n)-3n\gamma)}$$
After very simple mathematics we can recover other parameters which finally gives us possibility to perform a graphical analysis of $V(\phi)$ and $\omega_{\small{tot}}$ (Fig. \ref{fig:potintschange} and \ref{fig:omegatotschange}).
\begin{equation}\label{eq:TFcintrho}
\rho_{\small{TF}}=\frac{3n^{2}}{t^{2}}-\rho_{m}.
\end{equation}
\newpage
\section*{Discussion}
A mixture of Tachyonic dark energy and a fluid with a parameterized EoS was considered. EoS of the "new fluid" taken to be a function of a linear combination of Hubble parameter, power of Hubble parameter and its derivatives. From non interaction between two components up to 3 different types of interactions: $Q=3Hb\rho_{m}$, $Q=\alpha\dot{\rho}_{m}$ and recently proposed interaction called sign-changeable interaction $Q=q(\alpha\dot{\rho}_{m}+3Hb\rho_{m})$ involving deceleration parameter, was considered in this article. For all cases we are able to recover field $\phi$ and potential $V(\phi)$. Graphical analysis of $V$ evolution during time shows that we can recover real properties of tachyonic field: $V\rightarrow0$ with time. For some combination of the values of the parameters satisfying mentioned condition $\omega_{\small{tot}}$ also was investigated. Analysis shows that for all cases $\omega_{\small{tot}}>-1$ indicating quintessence-like behavior. By this article we would like to extend a part of \cite{Ujjal}, where one type of interaction: $Q=3Hb\rho_{m}$ was considered between field and a barotropic fluid: a special case of the fluid considered there.
\section*{Acknowledgments}
This research activity has been supported by EU fonds in the frame of the program FP7-Marie Curie Initial Training Network INDEX NO.289968.\\ \\

\newpage
\begin{figure}
\centering
\includegraphics[width=70 mm]{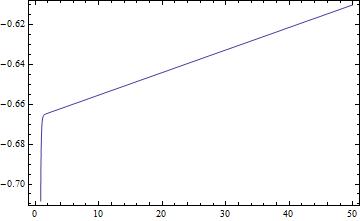}
\caption{%
\small{The variation of $\omega_{\small{tot}}$ against t, Interaction: $Q=b\dot{\rho}_{m}$,
Parameters:  $n=2$, $\rho_{0}=1$, $p_{0}=0$, $\gamma=2$, $\omega_{H}=1.5$, $\omega_{H2}=0.5$ and $\omega_{dH}=1$, $b=0.05$.}} 
\label{fig:omegatotdrho}
\end{figure}
\end{document}